\documentclass{aa}
\usepackage[dvips]{graphics}
\usepackage{latexsym}
\def\gtsima{$\; \buildrel > \over \sim \;$}
\def\ltsima{$\; \buildrel < \over \sim \;$}
\def\prosima{$\; \buildrel \propto \over \sim \;$}
\def\gsim{\lower.5ex\hbox{\gtsima}}
\def\lsim{\lower.5ex\hbox{\ltsima}}
\def\simgt{\lower.5ex\hbox{\gtsima}}
\def\simlt{\lower.5ex\hbox{\ltsima}}
\def\simpr{\lower.5ex\hbox{\prosima}}

\def\h1{$h^{-1}$}
\def\beq{\begin{equation}}
\def\eeq{\end{equation}}
\begin{document}
\title{
The K20 survey. III. Photometric and spectroscopic properties of the sample\thanks{Based 
on observations made at the European Southern Observatory,
Paranal, Chile (ESO LP 164.O-0560).}}
\author{
	A. Cimatti \inst{1}
	\and
	M. Mignoli \inst{2}
	\and
	E. Daddi \inst{3}
	\and
	L. Pozzetti \inst{2}
	\and
	A. Fontana \inst{4}
	\and
	P. Saracco \inst{5}
	\and
	F. Poli \inst{6}
	\and
	A. Renzini \inst{3}
	\and
	G. Zamorani \inst{2}
	\and
	T. Broadhurst \inst{7}
	\and
	S. Cristiani \inst{8,9}
	\and
	S. D'Odorico \inst{3}
	\and
	E. Giallongo \inst{4}
	\and
	R. Gilmozzi \inst{3}
	\and
	N. Menci \inst{4}
}
\institute{ 
Istituto Nazionale di Astrofisica, Osservatorio Astrofisico di Arcetri, Largo E. Fermi 5, I-50125 Firenze, Italy 
\and Istituto Nazionale di Astrofisica, Osservatorio Astronomico di Bologna, via Ranzani 1, I-40127, Bologna, Italy
\and European Southern Observatory, Karl-Schwarzschild-Str. 2, D-85748, 
Garching, Germany
\and Istituto Nazionale di Astrofisica, Osservatorio Astronomico di Roma, via Dell'Osservatorio 2, Monteporzio, 
Italy
\and Istituto Nazionale di Astrofisica, Osservatorio Astronomico di Brera, via E. Bianchi 46, Merate, Italy
\and Dipartimento di Astronomia, Universit\`a ``La Sapienza'', Roma, Italy
\and Racah Institute for Physics, The Hebrew University, Jerusalem, 91904, Israel
\and ST, European Coordinating Facility, Karl-Schwarzschild-Str. 2, D-85748, 
Garching, Germany
\and Istituto Nazionale di Astrofisica, Osservatorio Astronomico di
Trieste, Via Tiepolo 11, I-34131, Trieste, Italy
}
\offprints{Andrea Cimatti, \email{cimatti@arcetri.astro.it}}
\date{Received ; accepted }
\abstract{The K20 survey is an ESO VLT optical and near-infrared 
spectroscopic survey aimed at obtaining spectral information and 
redshifts of a complete sample of about 550 objects to $K_s\leq20.0$
over two independent fields with a total area of 52 arcmin$^2$. In this 
paper we discuss the scientific motivation of such a survey, we
describe the photometric and spectroscopic properties of the sample, 
and we release the $K_s$-band photometric catalog. Extensive 
simulations showed that the sample is photometrically highly complete 
to $K_s=20$. The observed galaxy counts and the $R-K_s$ color
distribution are consistent with literature results. 
We observed spectroscopically 94\% of the sample, reaching a spectroscopic 
redshift identification completeness of 92\% to $K_s\leq20.0$ for the 
observed targets, and of 87\% for the whole sample (i.e. counting also 
the unobserved targets). Deep spectroscopy was complemented with 
multi-band deep imaging in order to derive tested and 
reliable photometric redshifts for the galaxies lacking spectroscopic 
redshifts. The results show a very good agreement between the spectroscopic
and the photometric redshifts with $<z_{spe}-z_{phot}>=0.01$ and with
a dispersion of $\sigma_{\Delta z}$=0.09.
Using both the spectroscopic and the photometric redshifts, we reached 
an overall redshift completeness of about 98\%.
The size of the sample, the redshift completeness, the 
availability of high quality photometric redshifts and multicolor 
spectral energy distributions make the K20 survey database one of the 
most complete samples available to date for constraining the currently 
competing scenarios of galaxy formation and for a variety of 
other galaxy evolution studies.
}
\titlerunning{The K20 survey sample}
\authorrunning{A. Cimatti et al.}  \maketitle

\section{Introduction}

Contrary to surveys for high-$z$ galaxies selected in the optical, which 
are more sensitive to the star-formation activity, the selection of galaxies
in the $K$-band has the important advantage of not being
affected by strong K-correction effects. This is due to the similarity
of the different galaxy spectral types in the near-infrared.
Cowie et al. (1994) showed that over a broad range of redshifts
the K-correction amplitude in the $K$-band is much smaller than
those in $I$- or in $B$- bands, and very little dependent on the
galaxy spectral types (from elliptical to irregular galaxies).
Thus, the great advantage of a $K$-band selection is that the
resulting samples are almost free from strong selection effects and do
not critically depend on the galaxy types as in optical samples.

However, deriving the complete redshift distribution of a sample of faint
$K$-band selected galaxies is a challenging goal
even for 8-10m class telescopes. Indeed, a fraction
of the galaxies are beyond the spectroscopic limits (e.g.
very red galaxies with $R>26$) and/or lie in a redshift
range where the spectra do not present prominent features
in the observer frame (e.g. the redshift ``desert'' at
$1.4<z<2.0$ for optical spectroscopy).

Songaila et al. (1994) carried out an optical spectroscopic
survey of $K$-selected galaxies at different magnitude
depths over a wide range of field sizes (from $\sim$0.4 deg$^2$
down to $K$ \ltsima 15, to $\sim$5 arcmin$^2$ to $K$ \ltsima 20)
reaching an identification
completeness of $\sim$70\% at $K\sim 19-20$. Cowie et al. (1996)
performed with the Keck I 10m telescope a deeper optical spectroscopic
survey of galaxies with $K<20$ over a field of 26.2 arcmin$^2$,
reaching a spectroscopic identification completeness of $\sim$60\% at
$K\sim 20$. Cohen et al. (1999a, 1999b) observed a sample of 195
objects over an area of 14.6 arcmin$^2$ reaching a spectroscopic
completeness of 83.6\% to $K_s<20$. A large spectroscopic survey is
currently in progress over four independent fields covering a total
area of about 100 arcmin$^2$ to $K<20$ (Stern et al. 2001).

In addition to purely spectroscopic surveys, a number of more recent
projects made use mostly of photometric redshifts.
For instance, Drory et al. (2001) selected $\sim$5000 galaxies
to $K<19$ over 998 arcmin$^2$, and estimated the photometric redshifts
for 94\% of them using $VRIJK$ photometry (spectroscopic redshifts
were available for only $\sim$6\% of the sample).
Other surveys based largely on photometric redshifts range from
that of Firth et al. (2002), who covered a field of 744 arcmin$^2$
to $H<20.0-20.5$ and derived photometric redshifts for about
4000 galaxies using $UBVRIH$ photometry, to those of e.g. Fontana et al.
(1999,2000) and Rudnick et al. (2001) based on very deep observations
(typically to $K_s<21-23$) of very small fields (a few arcmin$^2$ ).

Surveys for faint galaxies selected in the $K$-band are very important
to investigate the formation and evolution of galaxies.

One of the main questions of galaxy evolution is whether massive 
galaxies (e.g. ${\cal M}_{stars}$\gtsima $10^{11}$ M$_{\odot}$) are the 
late product of merging of pre-existing disk galaxies occurring mostly
at $z<1.5-2$, as predicted by the current CDM scenarios (e.g. Kauffmann 
et al. 1993, 1996; Cole et al. 2000; Baugh et al. 1996; Somerville 
et al. 2001, Baugh et al. 2002, and references therein), or whether they 
formed at $z>2-3$ during a short-lived and single event of vigorous 
star formation, followed by a passive evolution (or pure luminosity 
evolution, PLE) of the stellar population to nowadays (see e.g. 
Renzini 1999; Renzini \& Cimatti 1999; Peebles 2002, for recent reviews).

Since the near-IR light is a good tracer of a galaxy stellar mass 
(Gavazzi et al. 1996; Madau, Pozzetti \& Dickinson 1998; Kauffmann \& 
Charlot 1998, KC98 hereafter), 
the above scenarios can in principle be tested by selecting galaxies 
at 2.2$\mu$m in the $K$-band and deriving their redshift distribution 
and physical 
and evolutionary properties (e.g. Broadhurst et al. 1992). 

KC98 estimated that $\sim 60\%$ and $\sim 10\%$ of the galaxies in 
a $K<20$ sample are expected to be at $z>1$, respectively, in a PLE 
and in a standard CDM hierarchical merging model (cf. their Fig. 4). 
Such a large difference was in fact one of the main motivations of 
our original project to undertake a complete redshift survey for all 
objects with $K<20$ in a small area of the sky. However, more recent 
models consistently show that for $z>1$ the difference between the 
predictions of different scenarios is less extreme than in the 
KC98 realization (Menci et al., Pozzetti et al., in preparation; see 
also Firth et al. 2002). Part of the effect is due to the now favored 
$\Lambda$CDM cosmology which pushes most of the merging activity in 
hierarchical models at earlier times compared to $\tau$CDM and 
SCDM models, and therefore get closer to the PLE case. Moreover, a 
different tuning of the star-formation algorithms (to accomodate for 
more star formation at high-$z$) also reduces the differences between 
the two scenarios (e.g. Somerville et al. 2001). 

In order to address the question of the formation and evolution of
galaxies and to constrain the current models, we performed a
new spectroscopic survey of $K$-selected galaxies with the ESO VLT.
In this paper, we present the main scientific motivations, we define 
the sample selection, we discuss the photometry, the completeness, 
we present the main spectroscopic and photometric redshift results,
and we release the $K_s$-band photometric catalog. The spectral and 
multicolor catalogs will be presented elsewhere together with the 
forthcoming papers on the scientific analysis of the sample.
Magnitudes are given in Vega system. The widely accepted 
cosmology with $H_0=70$ km s$^{-1}$ Mpc$^{-1}$, $\Omega_m=0.3$ 
and $\Omega_{\Lambda}=0.7$ is adopted throughout this paper.

\section{The K20 Survey} 

In order to overcome the limitations of the previous surveys
(e.g. small samples and/or spectroscopic incompleteness), we started 
in 1999 a project that was dubbed ``K20 survey''. To such 
project, 17 nights were allocated as an 
ESO VLT Large Program distributed over a period of two years 
(see also {\tt http://www.  arcetri.astro.it/$\sim$k20/} for
more details).

The survey aims at obtaining spectroscopic redshifts with 
the highest possible completeness for a sizeable sample of field 
galaxies with 
$K_s<20$. The main scientific aim of the K20 survey is to 
probe the evolution of galaxies to $z$ \gtsima 1,
deriving their redshift distribution, luminosity 
function and stellar mass function. This body of observational 
evidences is being and will be used for a comparison with
the different models of galaxy formation and evolution.

To this purpose, deep optical 
and near-infrared spectroscopy were complemented by deep multi-band 
imaging in order to derive tested and reliable photometric redshifts 
for the unavoidable fraction of galaxies with no spectroscopic redshifts. 

Besides such a main goal, the K20 survey was also designed
to address other important issues such as: (1) the nature of Extremely
Red Objects (EROs), (2) the evolution of elliptical galaxies, 
(3) the evolution of galaxy clustering, 
(4) the spectral properties of a large number of galaxies and their 
evolution as a function of redshift, (4) the evolution of the
volume star formation 
density, (5) the fraction of AGN in $K$-selected samples, (6) 
the optimization of the photometric redshift techniques based on 
ground-based imaging, and (7) the brown dwarf population at high 
Galactic latitude. The K20 database will also provide important
information to complement other multiwavelength surveys performed 
in the same fields, such as the SIRTF GOODS ({\tt http://www.stsci.
edu/science/goods}) and SWIRE ({\tt http://www.ipac.caltech.edu/SWIRE/})
projects.

Two papers based on the K20 survey have been already published showing 
an application of our database to investigate the nature and the 
clustering of the EROs (Cimatti et al. 2002; Daddi et al. 2002).

\begin{table*}
\caption[]{The $K_s$-band images of the target fields}
\begin{tabular}{lcccccccc} \hline\hline
& & & & & & & & \\
Field & Center Coordinates & Field Size & Seeing & Int. time & Gal. Coord. & $A_V$ & $A_K$ & \\
 & (J2000) & (arcmin$^2$) & (FWHM) & (hours) & & (mag) & (mag) & \\
& & & & & & & & \\
QSO 0055-269 & 00$^{h}$57$^{m}$58.91$^{s}$ -26$^{\circ}$43$^{\prime}$12.1$^{\prime\prime}$ & 19.8 & 0.9$^{\prime\prime}$ & 3.93 & 197.7$^{\circ}$ -88.5$^{\circ}$ & 0.065 & 0.007 \\
CDFS & 03$^{h}$32$^{m}$22.52$^{s}$ -27$^{\circ}$46$^{\prime}$23.5$^{\prime\prime}$ & 32.2 & 0.8$^{\prime\prime}$ & 3.0 & 223.5$^{\circ}$ -54.5$^{\circ}$ & 0.028 & 0.003 \\
& & & & & & & & \\ \hline\hline
\end{tabular}
\end{table*}

The sample was selected from two independent fields in order
to reduce the effects of the field-to-field variations. The 
targets were extracted from a sub-area of the Chandra Deep Field 
South (CDFS; Giacconi et al. 2001) and from a field centered 
around the QSO 0055-269 at $z$=3.656 (see Tab. 1 for details). 
The total sample includes 546 objects over 52 
arcmin$^{2}$ selected on the basis of the single criterion of
$K_s\leq 20.0$. 
Adopting the K-correction effects derived from the Bruzual
\& Charlot (1993) spectral synthesis models (GISSEL version 2000, 
hereafter BC2000), an apparent magnitude of $K_s=20.0$ corresponds 
to rest-frame typical absolute magnitudes of $M_{K_s}\sim$-21.7 and 
$M_{K_s}\sim$ -23.4 for $z=0.5$ and $z=1.0$ respectively, with a weak 
dependence on the galaxy type because of the small K-correction effects. 
Such absolute magnitudes correspond to $\sim0.1 L_K^*$ and $\sim0.4 L_K^*$ 
respectively 
(adopting the $L_K^*$ of the Cole et al. 2001 near-IR local luminosity 
function of galaxies). Since the stellar mass-to-infrared luminosity ratio 
(${\cal M}_{stars}/L_K$) is relatively insensitive to the star 
formation history, the limiting $K_s$-band luminosities of our survey 
can be converted into the corresponding limiting stellar masses. As 
the stellar population ages, ${\cal M}_{stars}/L_K$ (in solar units) 
remains very close to unity, independent of the galaxy color and 
Hubble type, and has a weak dependence on $z$. 

According to the mean stellar mass-to-light ratio and representative
stellar mass ${\cal M}^*_{stars}$ in the local universe (Cole et al.
2001), and adopting the predictions of the BC2000
spectral synthesis models, the limiting stellar masses corresponding
to $K_s$=20, for the Salpeter IMF, are about ${\cal M}_{stars}>10^{10} 
M_{\odot}\simeq0.07 {\cal M}^*_{stars}$ and ${\cal M}_{stars}>4\times10^{10}
M_{\odot}\simeq0.3 {\cal M}^*_{stars}$ for $z=0.5$ and $z=1$ respectively.

\section{Near-infrared imaging}

For both fields, the $K_s$-band images were obtained with the ESO NTT
equipped with SOFI (Moorwood et al. 1998) with a pixel size of
0.29$^{\prime\prime}$ and under photometric conditions (see also 
Rengelink et al. 1998 for more details on observations and database 
for the CDFS).

Since the ESO Imaging Survey (EIS; {\tt http://www.eso.org/science/eis/}) 
public $JK_s$ images of the CDFS available at the time of the K20 sample 
selection were affected by a loss of the flux up to 0.3 magnitudes at 
the faint limit of our survey, we performed a new independent 
reduction and calibration of the CDFS $K_s$-band image. 

The $K_s$-band data of the CDFS and the $JK_s$ data of the 0055-2659
field were reduced in a standard manner using the IRAF\footnote{
IRAF is distributed by the National Optical Astronomy Observatories, 
which are operated by the Association of Universities for Research 
in Astronomy, Inc., under cooperative agreement with the National 
Science Foundation.} software package DIMSUM 
\footnote{Deep Infrared Mosaicing Software, a package written 
by Eisenhardt, Dickinson, Stanford and Ward, available at 
ftp://iraf.noao.edu/contrib/dimsumV2.}. Raw frames were first 
corrected for bias and dark current by subtracting a median 
dark frame, and flat-fielded using an average differential sky 
light flat-field image. After deriving a master mask frame
with DIMSUM, a sky background image for each frame was obtained
by averaging a set (from 6 to 10) of time adjacent frames where
the sources were masked out. The sky-subtracted frames were then
inspected in order to reject or to correct the frames with bad sky
residuals, and the flux of bright objects visible in the individual 
frames was monitored to verify the photometric conditions and to
reject discrepant frames if necessary. The final sky-subtracted 
frames were rescaled to the same airmass and photometric zero-point, 
shifted and coadded together to generate the final image. The 
photometric calibration was achieved through the observation of 
standard stars 
from the Persson et al. (1998) sample. The Galactic extinction was 
estimated using the maps of Schlegel et al. (1998) (see Tab. 1). 

\section{Photometry and completeness of the sample}

\subsection{Source extraction and photometry}

The sample was extracted from the $K_s$-band images using the 
SExtractor package (Bertin \& Arnouts 1996). After convolving the images
with a Gaussian function matching the measured seeing FWHM, the objects
exceeding a $S/N>2.2$ over the background noise in a minimum area of 10
pixels (i.e. 0.7$\sigma$/pixel) were extracted. 

The total flux of the selected objects was then measured using the 
SExtractor BEST magnitude. Such a magnitude is defined to be either
a Kron magnitude (Kron 1980), which is measured in an elliptical
aperture whose size is determined by the profile of the object, or
an isophotal magnitude for objects blended or in crowded fields
(see Bertin \& Arnouts 1996 for more details). 

In both fields, the images are considerably deeper than the $K_s=20$ 
threshold of the spectroscopic sample, thus minimizing both selections 
effects and biases in the estimated magnitudes. Typically, galaxies 
at the faint limit of our sample have $S/N\simeq 10-15$, according 
to the SExtractor noise estimate. The cases of blended objects 
were analysed in more detail in order to correct the total magnitudes 
for the effects of blending.

\subsection{Photometric completeness}

Since the goal of photometric surveys for faint galaxies is
to measure their total integrated flux, the completeness is usually 
defined only by the total apparent magnitude. However, in practice 
it is well known that a fraction of the flux is lost when measuring 
magnitudes with aperture or isophotal photometry because of the
finite size of the photometric apertures and a minimum 
area for object detection (e.g. Impey \& Bothun 1997; Dalcanton 1998). 
The fraction of the lost light at the survey limiting flux
depends on the size and on the surface brightness profile of the
galaxies. Moreover, since the surface brightness of galaxies rapidly 
becomes fainter with increasing redshift as $(1+z)^{-4}$, this 
cosmological dimming could make many high-$z$ galaxies undetectable
below the adopted surface brightness threshold.

Thus, it is important to evaluate such potential selection effects
in order to make a meaningful comparison between the observed properties 
of the sample and the model predictions of galaxy formation and evolution.
We have therefore explored in our survey the effects due to the detection 
limits, i.e. surface brightness limit ($\mu_K \sim 23$ mag/arcsec$^2$) 
and minimum area (10 pixels), and to the procedure adopted for
the photometric measurements,
using both real and artificial galaxies.

\subsubsection{Limiting surface brightness}

Following Lilly et al. (1995), we show in Figure 1 the surface brightness 
at the minimum detection radius, 
as a function of magnitude for all the objects in our survey.
The vertical line marks the magnitude limit, $K_s<20$, while the horizontal
lines show the surface brightness limits used in the objects detection
for the two fields (i.e. a source is detected only if its detection surface 
brightness is above the surface brightness limit).
Most of the objects have surface brightness well above the nominal limit.

Figure 1 also shows the tracks expected for $L^*$ and $10L^*$ elliptical
and spiral galaxies. These tracks have been obtained using the observed
relations between luminosity, effective radius and surface brightness
for local spirals (Impey et al. 1996) and elliptical (Bender et al. 
1992, Pahre 1999) galaxies, and taking into account the
the effect of the seeing on the intrinsic galaxy profile (see Angeretti, 
Pozzetti \& Zamorani 2002 for details). The adopted galaxy profile 
parameters are indicated in the Figure caption. Note that at the faint 
limit of the sample, the L$^*$ tracks are essentially parallel to the 
star locus because for these galaxies the observed profiles are
dominated by the seeing. Fig. 1 shows that the surface 
brightness selection effects are negligible for $L<10 L^*$ galaxies. 

\begin{figure}[ht]
\resizebox{\hsize}{!}{\includegraphics{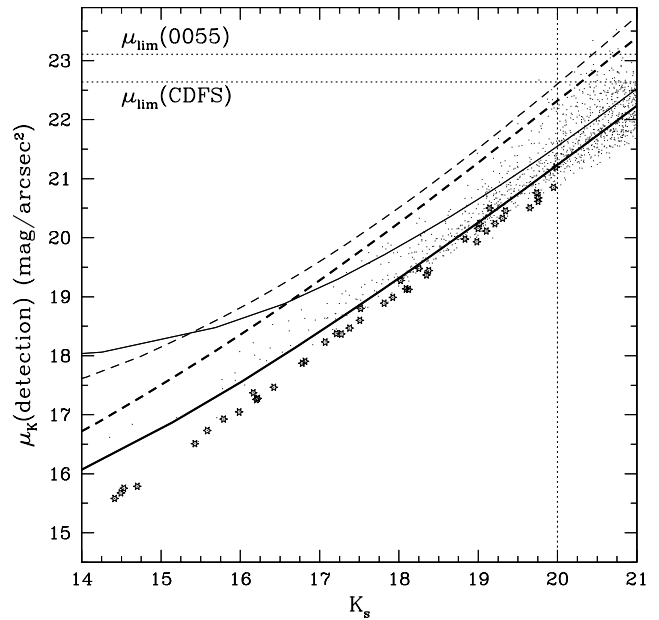}}
\caption{\footnotesize
Surface brightness at the minimum detection radius as a function of 
magnitude for all
objects in the sample. The vertical dotted line represents the
sample magnitude limit, $K_s<20$, while the horizontal dotted
lines indicate the
limiting detection surface brightness in the two fields
($\mu_K\sim22.64,
23.11$ mag/arcsec$^2$ in the CDFS and 0055-269 field respectively).
The tracks indicate the expected behavior of elliptical  (thick lines)
and spiral (thin lines) galaxies for two representative luminosities
(Kochanek et al. 2001):
L$^*$ (solid lines: $M^*_K=-24.3+5$log $h_{70}$, $r_{e}=3.1
~h_{70}^{-1}$ kpc
for ellipticals and $M^*_K=-23.8+5$log $h_{70}$, $r_{e}=6.1
~h_{70}^{-1}$ kpc
for spirals) and 10 L$^*$ (dashed lines: $M^*_K=-26.8+5$log $h_{70}$,
$r_{e}=16.3 ~h_{70}^{-1}$ for ellipticals and $M^*_K=-26.3+
5$log$h_{70}$,
$r_{e}=13.0 ~h_{70}^{-1}$ for spirals).
Star symbols represent the stars identified in our survey on the basis
of
their spectra.
}
\label{fig:plot}
\end{figure}

\begin{figure}[ht]
\resizebox{\hsize}{!}{\includegraphics{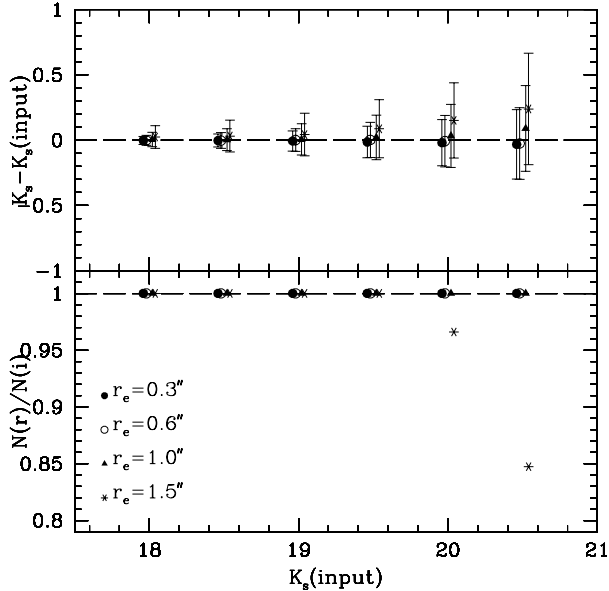}}
\caption{\footnotesize 
The results of a simulation in the 0055-269 field made
with artificial galaxies with de Vaucouleurs profiles, $b/a$=0.7,
four values of $r_e$ and for 6 input magnitudes ($K_s$=18.0,18.5,19.0,
19.5,20.0,20.5) (see text for more details). The top and the bottom 
panels show respectively the average difference between the measured ($K_s$) 
and the input magnitudes, and the fraction of the number of recovered 
objects, N(r), to the number of input artificial objects, N(i). The 
points are shifted along the abscissa around each of the 6 magnitudes 
to improve their visibility. Similar results are obtained for the CDFS.
}
\label{fig:plot}
\end{figure}

\begin{figure}[ht]
\resizebox{\hsize}{!}{\includegraphics{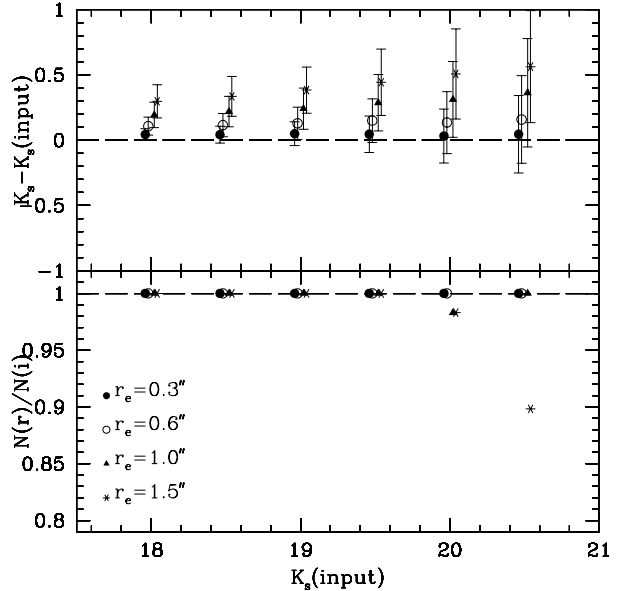}}
\caption{\footnotesize 
Same as Fig. 2, but for a simulation in the 0055-269 field made
with exponential profiles and $b/a$=0.4 Similar results are obtained 
for the CDFS.
}
\label{fig:plot}
\end{figure}

\subsubsection{Adding artificial galaxies}

In order to better assess the effects for different galaxy types,
we applied a method consisting in the addition of artificial
galaxies to the real images of our survey fields.

About 100 artificial galaxies were generated using the 
{\tt artdata.mkobjects} software in IRAF. We used three morphological 
types (one de Vaucouleurs profile with axial ratio $b/a$=0.7, and two 
exponential profiles with $b/a$=0.4 and $b/a$=0.8 respectively), 
and four intrinsic half-light radii for each morphology type ($r_e$=0.3
$^{\prime\prime}$,0.6$^{\prime\prime}$,1.0$^{\prime\prime}$,1.5
$^{\prime\prime}$). The choice of such values for $r_e$ was
motivated by the quantitative morphological studies of faint galaxies 
which showed that objects with $r_e$\gtsima $1.0^{\prime \prime}$ are 
extremely rare (e.g. Marleau \& Simard 1998). 

Such 12 artificial galaxy models were then 
convolved with a PSF with a Moffat profile (as derived from 
field stars) and created for different magnitudes ($K_s=18.0,
18.5,19.0,19.5,20.0,20.5$). In both fields, the results indicate
that at $K_s=20.0$ the completeness ($N_{recovered}/N_{input}$) is 
100\% for $r_e\leq1.0^{\prime\prime}$, whereas it decreases to 
$\sim$90\% for $r_e=1.5^{\prime\prime}$. We also verified that the 
results do not significantly change for other values of $b/a$. 

Fig. 2 and Fig. 3 show examples of the results of such simulation.
The recovered magnitudes at $K_s\leq20.0$ are generally consistent 
with the input ones for the exponential profile galaxies, but 
they turned out to be systematically underestimated for galaxies 
with a de Vaucouleurs profile (typically 0.1-0.3 magnitudes for 
$0.3^{\prime\prime}\leq r_e \leq 1.0^{\prime\prime}$). 

Such an effect was already discussed for example by Fasano et 
al. (1998), Dalcanton (1998) and Martini (2001). The origin of 
such an effect is due to the magnitudes being limited to an 
aperture which exclude the outermost regions of the galaxy 
surface brightness profiles, and it does not depend on the 
software used for the automatic photometry.

While this is less critical for rapidly declining exponential
profiles, its influence is larger for the more slowly decreasing 
de Vaucouleurs profiles. Being typically $r_e$\ltsima $0.6^
{\prime\prime}$ in high-$z$ early-type galaxies (e.g. Fasano et 
al. 1998), this means that the underestimate 
in our sample is at most \ltsima 0.2 magnitudes at $K_s=20$ 
for galaxies with a de Vaucouleurs profile. The lack of HST 
imaging for the K20 sample prevented us from applying magnitude 
corrections for the galaxies with de Vaucouleurs 
profiles. 

\begin{figure}[ht]
\resizebox{\hsize}{!}{\includegraphics{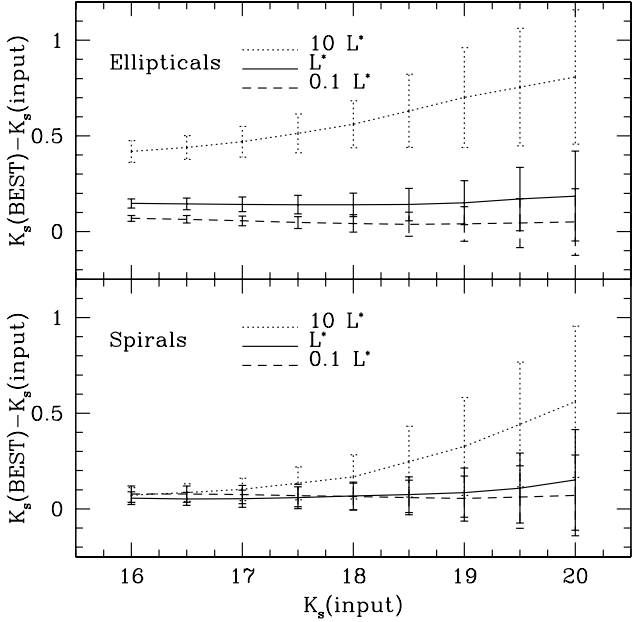}}
\caption{\footnotesize 
A comparison of BEST magnitude from SExtractor
and input magnitudes in our simulations for ellipticals (top panel)
and spirals (bottom panel) in the CDFS. Different lines represent different
luminosities, as indicated in the labels, and the error bars represent
the dispersion in the simulations, using about 100 artificial galaxies
at each magnitudes.
}
\label{fig:plot}
\end{figure}

Figure 4 shows the same results as a function of the luminosity 
(adopting the luminosity-surface brigthness relations discussed in
section 4.2.1). 
For typical L$^*$ ellipticals we could miss systematically
up to 20\% of the flux, and only $<$10-15\% for exponential disks. Such 
effects become more relevant only for very large and high luminosity galaxies
(e.g. $L\sim 10 L^*$), but fortunately such objects are rare enough
not to introduce severe selection effects in our survey. 

Such selection effects will be taken into account in the next papers 
including the comparison with the predictions of galaxy formation models.

\subsubsection{Dimming real galaxies}

As an additional test on the completeness, we applied the method 
described by Saracco 
et al. (1999, 2001), consisting in dimming observed bright galaxies by 
various factors, while keeping the 
background noise constant and equal to that of the original image.
In particular, all the objects were extracted from the original frame,
kept with the same original sizes, scaled to a lower flux and 
then re-inserted in the real images. 
A new catalog has then been obtained 
on these ``dimmed'' images with the same parameters used for the 
original one. This method tests the whole photometric procedure 
using a set of objects with clustering and with a wide variety of
morphological properties representative of the real galaxy population. 

These simulations have shown that: {\it a)} the flux fraction lost by 
the BEST magnitudes is negligible to about $K_s=19.5$, whereas
a slight underestimate of 0.05-0.1 magnitudes is present in the 
fainter side of the sample $19.5<K_s<20.0$; {\it b)} 100\% of the
objects 
at $K_s=20$ are robustly detected in both fields, irrespective of their
morphology, although {\it c)} the scatter between the input and the 
recovered magnitudes is typically 0.1 magnitudes at $K_s>19.5$. 
As a result, {\it d)} the number counts observed in both fields are 
not affected by significant instrumental incompleteness.

\begin{table}
\caption[]{The differential galaxy counts}
\begin{tabular}{crrrr} \hline\hline
& & & & \\
$K_s$ & $n_{0055-269}$ & $n_{CDFS}$ & $n_{tot}$ & $N_{tot}$ \\
& & & &  \\
13.75 & 0$^{+1.84}_{-0.00}$ & 1$^{+2.30}_{-0.83}$  & 1$^{+2.30}_{-0.83}$ & 69.2$^{+159.2}_{-57.5}$ \\
14.25 & 0$^{+1.84}_{-0.00}$ & 1$^{+2.30}_{-0.83}$  & 1$^{+2.30}_{-0.83}$ & 69.2$^{+159.2}_{-57.5}$ \\
14.75 & 1$^{+2.30}_{-0.83}$ & 0$^{+1.84}_{-0.00}$  & 1$^{+2.30}_{-0.83}$ & 69.2$^{+159.2}_{-57.5}$ \\
15.25 & 0$^{+1.84}_{-0.00}$ & 1$^{+2.30}_{-0.83}$  & 1$^{+2.30}_{-0.83}$ & 69.2$^{+159.2}_{-57.5}$ \\
15.75 & 2$^{+2.64}_{-1.29}$ & 3$^{+2.92}_{-2.37}$  & 5$^{+3.38}_{-2.16}$ & 346.1$^{+234.0}_{-149.5}$ \\
16.25 & 1$^{+2.30}_{-0.83}$ & 5$^{+3.38}_{-2.16}$  & 6$^{+3.58}_{-2.38}$ & 415.4$^{+247.8}_{-164.8}$\\
16.75 & 3$^{+2.92}_{-2.37}$ & 5$^{+3.38}_{-2.16}$  & 8$^{+3.95}_{-2.77}$ & 553.8$^{+273.5}_{-191.8}$\\
17.25 & 7$^{+3.77}_{-2.58}$ & 17$^{+5.20}_{-4.08}$ & 24$^{+5.97}_{-4.86}$ & 1661.5$^{+413.3}_{-336.5}$ \\
17.75 & 15$^{+4.96}_{-3.83}$ & 22$^{+5.76}_{-4.65}$ & 37$^{+7.14}_{-6.06}$ & 2561.5$^{+494.3}_{-419.5}$ \\
18.25 & 28$^{+6.35}_{-5.26}$ & 35$^{+6.97}_{-5.89}$ & 63$^{+8.00}_{-7.91}$ & 4361.5$^{+553.8}_{-547.6}$ \\
18.75 & 29$^{+6.45}_{-5.35}$ & 59$^{+7.75}_{-7.66}$ & 88$^{+9.43}_{-9.36}$ & 6092.3$^{+652.8}_{-648.0}$ \\
19.25 & 56$^{+7.54}_{-7.46}$ & 69$^{+8.36}_{-8.28}$ &125$^{+11.18}_{-11.18}$ & 8653.8$^{+774.0}_{-774.0}$ \\
19.75 & 43$^{+7.61}_{-6.53}$ & 86$^{+9.33}_{-9.25}$ &129$^{+11.36}_{-11.36}$ & 8930.7$^{+786.5}_{-786.5}$ \\
& & & & \\ \hline\hline
\end{tabular}

\small
$K_s$: bin central magnitude; 

$n_{0055}$: number of galaxies in the 0055-269 field with 1$\sigma$
poissonian uncertainties; 

$n_{CDFS}$: number of galaxies in the CDFS with 1$\sigma$ poissonian
uncertainties; 

$n_{tot}$: total number of galaxies with 1$\sigma$ poissonian
uncertainties; 

$N_{tot}$: gal/deg$^2$/0.5 mag (both fields) with 1$\sigma$ poissonian
uncertainties; 

\normalsize
\end{table}

\subsection{Galaxy counts and colors}

Tab. 2 lists the galaxy differential counts in 0.5 magnitude bins in 
our survey, and Fig. 5 shows a comparison with a compilation of 
literature counts. No corrections for incompleteness were applied to our data,
and we excluded the AGNs and the stars with spectroscopic identification
in our sample. The 1$\sigma$ poissonian uncertainties of the counts were 
estimated according to the prescriptions of Gehrels (1986).
Our counts are in very good agreement with those of previous surveys. 

On the basis of what discussed in previous sections, the slight
flattening shown by our data for $K_s>19.5$ should not be due to
incompleteness, but is likely to be due to field-to-field fluctuations 
which, as Figure 5 shows, are the dominant source of scattering
in the counts from survey to survey. This is supported by the fact that
the small drop in our total counts to $K_s>19.5$ is seen in only 
one of our two fields (the 0055-269 field) (see Fig. 5).

\begin{figure}[ht]
\resizebox{\hsize}{!}{\includegraphics{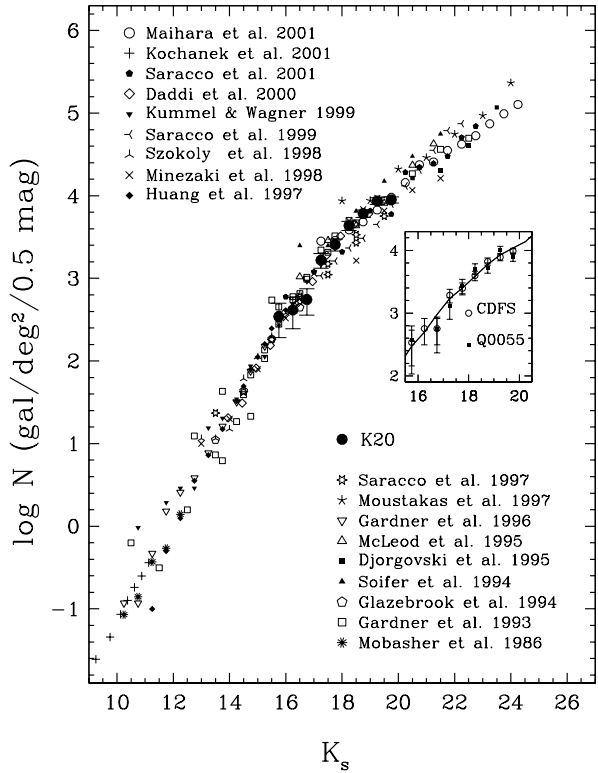}}
\caption{\footnotesize 
Differential $K_s$-band galaxy counts. The filled circles represent 
the K20 survey counts. Literature survey counts are shown with 
different symbols. The insert plot shows the counts separately
for the CDFS and the 0055-269 field, and the continuous line represents
the average $K_s$-band galaxy counts taken from Hall et al. (1998).
}
\label{fig:plot}
\end{figure}

The detailed color properties of the galaxies in the K20 
sample will be discussed elsewhere. Here, as an additional check 
of the photometry, we show a comparison between the $R-K_s$ galaxy 
colors of the galaxies in the K20 sample (the same color used to 
select the K20 ERO sample discussed in Cimatti et al. 2002 and 
Daddi et al. 2002) and the sample taken from the Caltech Faint Galaxy 
Redshift Survey (CFGRS; Cohen et al. 1999a, 1999b), a spectroscopic 
survey of a complete sample of $K_s<20.0$ galaxies over a field of 14.6 
arcmin$^2$. 

Fig. 6 shows a comparison between the $R-K_s$ color distributions. 
As suggested by Fig. 6 (top panel), such a comparison may be biased 
by the presence of three clusters (or rich groups) of galaxies present in 
both surveys: one at $z=0.67$ and one at $z=0.73$ in the 0055-269 and 
in the CDFS respectively, and one at $z=0.58$ in the CFGRS.  
To investigate this possibility, we excluded the members of the 
richest clusters in both samples (Fig. 6 bottom panel). As Figure 6 
(bottom panel) clearly suggests, a Kolmogorov-Smirnov test indicates 
that the hypothesis that the two distributions are drawn from the same 
population cannot be rejected at a confidence level of 0.05, thus 
demonstrating that the difference (Fig. 6, top panel) was due to 
color biases introduced by the galaxy clusters (or groups). 

\begin{figure}[ht]
\resizebox{\hsize}{!}{\includegraphics{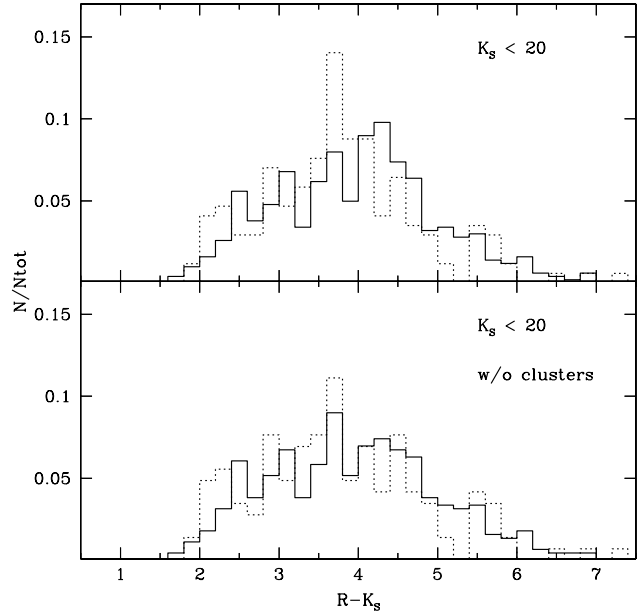}}
\caption{\footnotesize 
Top panel: the K20 and the CFGRS (Cohen et al. 1999a, 1999b) 
survey $R-K_s$ color distributions for $K_s<20.0$ galaxies 
(continuous and dotted lines respectively). Bottom panel: the
comparison between the K20 and the CFGRS $R-K_s$ color 
distributions after removing the richest clusters in the two samples.
}
\label{fig:plot}
\end{figure}

\section{Optical spectroscopy}

Optical multi-object spectroscopy was obtained with the ESO VLT UT1 and UT2
equipped respectively with FORS1 (with MOS mode: 19 movable slitlets) and 
FORS2 (with MXU mode: up to 52 targets observed through a laser-cut 
slit mask) during 0.5$^{\prime\prime}$-2.0$^{\prime\prime}$ seeing conditions and
with 0.7$^{\prime\prime}$-1.2$^{\prime\prime}$ wide slits depending on
the seeing. The grisms 150I, 200I, 300I were used, providing dispersions
of 5.5, 3.9, 2.6~\AA/pixel and spectral resolutions of R=260, 380, 660 respectively.
The integration times ranged from 0.5 to 3 hours. Dithering of the targets 
along the slits between two fixed positions was applied as much as possible 
for the faint target observations in order to efficiently remove the 
CCD fringing and the strong OH sky lines at $\lambda_{obs}>7000$~\AA. 

The data reduction was carried out using IRAF packages. The bias subtraction,
overscan correction and flat-fielding were performed in the standard mode.
For the non-dithered spectra, the sky background was removed from the 
two-dimensional frames obtained by fitting a third-order polynomial 
along the spatial direction and avoiding the regions with object spectra. 
The resulting sky-subtracted images were kept for following analysis in the 
course of the redshift determination. The 1D spectra were then optimally 
extracted and wavelength calibrated using exposures of He-Hg-Ar lamps, 
taken with the same instrumental configuration. Finally, the spectra were 
flux-calibrated using standard spectrophotometric stars, de-reddened for 
atmospheric extinction, corrected for telluric absorptions and scaled 
to the total $R$-band magnitudes.

The spectroscopic analysis of the optical spectra and the redshift estimates
were done using IRAF tasks, both automatic ({\tt xcsao}) and interactive
ones ({\tt rvidlines} and {\tt splot}), and always through visual inspection of
the 1D and 2D spectra. 

We followed different approaches in the redshift measurements, according 
to the quality and 
type of the analyzed spectrum. In case of a spectrum with emission 
lines, the redshift was obtained with {\tt rvidlines} by means of multiple 
Gaussian fitting of the spectral features. For the objects with fairly 
good signal--to--noise ratio and with an absorption line dominated spectrum, 
we performed an automatic cross-correlation with {\tt xcsao}. 
As templates for the cross-correlation we used the set of Kinney et al.
(1996) template spectra and also high $S/N$ composite spectra obtained
from our own K20 survey database. For objects with both emission and 
absorption lines, 
we used both {\tt rvidlines} and {\tt xcsao}. For spectra with
poor $S/N$ ratio, 
with a single emission line and/or uncertain spectral features,
the redshift was estimated both by checking the reality of
the structures on the 2D frames and interactively measuring the 
wavelengths of the features using {\tt splot} and cross-checking
the results with {\tt rvidlines}. Several and independent 
cross-checks on ambiguous cases were done independently by 2-3 of us  
in order to ensure independent redshift estimates and to eliminate 
discrepant cases. Objects with multiple observations made in different
mask configurations or different runs provided always consistent
results in terms of detected features and redshift measurements, and
their spectra were then coadded to increase the final signal-to-noise
ratio.

The spectroscopic redshifts were divided into two categories: 
``high-quality and secure'' redshifts (when several features 
were detected at high significance), and ``lower-quality'' redshifts 
(when only one weak emission line was detected and ascribed to 
[OII]$\lambda$3727 emission, or when the detected spectral features 
were more marginal). Objects with ``lower-quality'' redshifts are 5\% 
of the total sample to $K_s\leq20.0$, whereas their fractions
decrease to 4\% and 2\% for $K_s\leq19.5$ and $K_s\leq19.0$ respectively. 
One of the following spectral classes was preliminarly assigned to 
each object: star, Type 1 (i.e. with broad lines) AGN, emission line 
galaxy, early-type galaxy with 
emission lines, early-type galaxy with no emission lines. Fig. 7
shows examples of the different spectral types. A more detailed
spectral classification is under way and will be presented 
together with the K20 spectroscopic catalog.

\begin{figure}[ht]
\resizebox{\hsize}{!}{\includegraphics{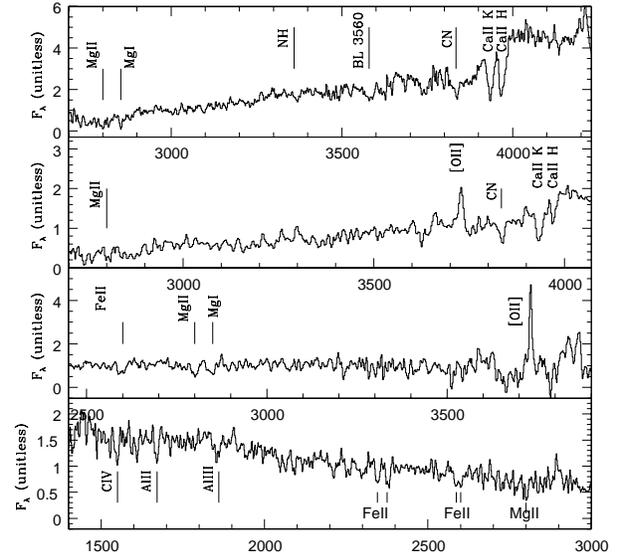}}
\caption{\footnotesize 
Examples of high-$z$ galaxies with the four adopted spectral classifications.
From top to bottom: an early-type at $z=1.096$ ($R=23.0$), an early-type
+ [OII]$\lambda$3727 emission at $z=0.735$ ($R=23$), an emission 
line galaxy at $z=1.367$ ($R=23.0$), an absorption line galaxy at 
$z=1.725$ ($R=23.5$). Some typical absorption and emission lines are 
also indicated for each spectrum.
}
\label{fig:plot}
\end{figure}

\begin{table*}
\caption[]{The VLT spectroscopic runs}
\begin{tabular}{lllcrlr} \hline\hline
& & & & & & \\
Dates & Telescope & Instrument & Mode & Multiplex & Seeing range & Run completion \\
& & & & & & \\
1999 01-05.10 & UT1 & FORS1 & MOS & $\leq$19 & 1.0$^{\prime\prime}$-2.5$^{\prime\prime}$ & 50\%\\
1999 06-09.11 & UT1 & FORS1 & MOS & $\leq$19 & 0.4$^{\prime\prime}$-1.5$^{\prime\prime}$ & 100\%\\
1999 28-30.11 & UT1 & ISAAC & LR & $\leq$2 & 0.5$^{\prime\prime}$-1.5$^{\prime\prime}$ & 100\%\\
2000 24-28.11 & UT2 & FORS2 & MXU & $\leq$52 & 0.4$^{\prime\prime}$-1.5$^{\prime\prime}$ & 100\%\\
2000 04-08.12 & UT1 & ISAAC & LR & $\leq$2 & 0.5$^{\prime\prime}$-2.0$^{\prime\prime}$ & 50\%\\
& & & & & & \\ \hline\hline
\end{tabular}
\end{table*}

\section{Near-infrared spectroscopy}

A small fraction (22 objects, 3.9\%) of the K20 sample was 
observed with near-IR spectroscopy 
using the VLT UT1+ISAAC (Moorwood et al. 1999) in order to derive the 
redshifts of the 
galaxies which were too faint for optical spectroscopy and/or 
expected to be in a redshift range difficult for optical spectroscopy.  

During the first ISAAC run (Table 3), targets in the CDFS were selected
according to the photometric redshifts available at that time
in order to search mostly for H$\alpha$ emission redshifted in the
$H$-band for $1.3<z<1.8$. The observations were carried out 
under photometric and 0.5$^{\prime\prime}$-1.5$^{\prime\prime}$ seeing
conditions. The low spectral resolution mode (R=500 with
1$^{\prime\prime}$ slit) was adopted in order
to maximize the covered spectral range per integration, the slit 
was 1$^{\prime\prime}$ wide, providing a spectral resolution FWHM of 
32~\AA~ in $H$-band. The observations were done by nodding the target(s)
along the slit between two positions A and B with a typical nod throw 
of 10-20$^{\prime\prime}$. The integration time was 10 minutes per
position. For instance, a total integration time of 1 hour was obtained 
following a pattern ABBAAB. Whenever possible, two targets were observed 
simultaneously in the slit in order to maximize the multiplex. The 
spectral frames were flat-fielded, 
rectified, sky-subtracted, coadded and divided by the response curve 
obtained using the spectra of O stars. Absolute flux calibration was 
achieved by normalizing the spectra to the $JK_s$ broad-band photometry
or to the interpolated $H$-band magnitude in case of $H$-band spectra. 

During the second ISAAC run (Tab. 3), the targets were selected
from the CDFS and the 0055-269 fields whenever their 
spectra did not show features in the optical and if the photometric
redshifts suggested $z>1.3$. The observations were carried out
as for the first run. Despite a large fraction of the time lost 
because of bad weather and poor seeing, such a run was more 
successful in identifying a few high-$z$ star-forming galaxies
thanks to their H$\alpha$ emission.

The bottom line for the ISAAC runs is that the absence of a 
multi-object spectroscopic mode, being each integration limited
to a single band ($J$ or $H$ or $K$), and the lack of very accurate 
photometric redshifts at the time of the observations made it 
difficult to derive a substantial number of spectroscopic redshifts based on 
near-IR spectroscopy alone. All in all, redshifts were derived for
four galaxies at $1.3<z<1.9$ out of the 22 observed (see Fig. 8 for
two examples of successful ISAAC observations).

\begin{figure}[ht]
\resizebox{\hsize}{!}{\includegraphics{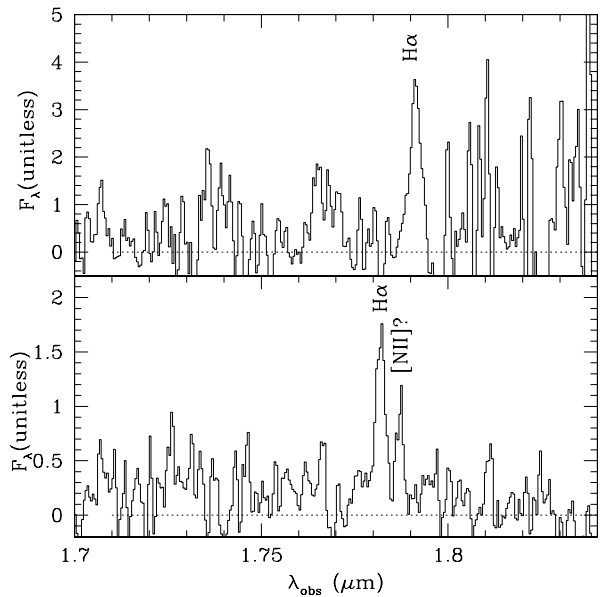}}
\caption{\footnotesize 
ISAAC $H$-band low resolution spectra of two emission line 
galaxies with H$\alpha$ redshifted at $z=1.729$ (top panel)
and $z=1.715$ (bottom panel).
}
\label{fig:plot}
\end{figure}

\begin{table*}
\caption[]{Summary of the spectroscopic results}
\begin{tabular}{llcccccccc} 
\hline
& & & & & & & & \\ 
Magnitude & Field & $N_{total}$ & $N_{observed}$ & $N_{identified}$ & $\frac{N_{identified}}{N_{total}}$ & $\frac{N_{identified}}{N_{observed}}$ & $N_{galaxies}$ & $N_{AGN}$ & $N_{stars}$ \\
& & & & & & & & \\ 
& & & & & & & & \\
\hline \\
$K_s\leq$19.0 & 0055-269 & 97  & 95  &  92 & 0.95 & 0.97 & 81  & 4 &   7 \\ \\ 
     & CDFS & 180 & 176 & 168 & 0.93 & 0.95 & 137 & 5 &  26 \\ \\ 
     & TOTAL & 277 & 271 & 260 & 0.94 & 0.96 & 218 & 9 & 33 \\ \\ 
\hline \\
$19.0<K_s\leq$19.5 & 0055-269 & 58 & 56 & 49 & 0.84 & 0.88 & 47 & 0 & 2 \\ \\ 
     & CDFS & 75 & 73 & 69 & 0.92 & 0.95 & 63 & 2 & 4 \\ \\ 
     & TOTAL & 133 & 129 & 118 & 0.85 & 0.91 & 118 & 2 & 6 \\ \\ 
\hline \\
$19.5<K_s\leq$20.0 & 0055-269 & 43 & 31 & 26 & 0.60 & 0.84 & 26 & 0 & 0 \\ \\ 
     & CDFS & 93 & 82 & 70 & 0.75 & 0.85 & 63 & 1 & 6 \\ \\ 
     & TOTAL & 136 & 113 & 96 & 0.71 & 0.85 & 89 & 1 & 6 \\ \\ 
\hline \\
$K_s\leq$20.0 & 0055-269 & 198 & 182 & 167 & 0.84 & 0.92 & 154 & 4 & 9 \\ \\
     & CDFS & 348 & 331 & 307 & 0.88 & 0.93 & 263 & 8 & 36 \\ \\
     & TOTAL & 546 & 513 & 474 & 0.87 & 0.92 & 417 & 12 & 45 \\ \\
& & & & & & & & \\
\hline
\end{tabular}
\end{table*}

\section{Spectroscopic completeness}

Table 4 summarizes the status of the sample and of the spectroscopic
redshift identifications. 

Only about 6\% of the sample targets to $K_s=20$ were not observed 
either because it was not possible to fit them in the observing masks 
or because too faint in the optical bands or because of the nights
lost due to bad weather or bad seeing conditions.

The efficiencies in deriving the spectroscopic redshifts 
($N_{identified}/N_{observed}$) was 96\%, 91\% and 85\% for 
$K_s\leq 19.0$, $19.0< K_s \leq 19.5$ and $19.5< K_s \leq 20.0$
respectively, whereas the overall spectroscopic redshift 
completeness ($N_{identified}/N_{total}$) in the same magnitude
ranges is 94\%, 85\% and 71\%.

The spectroscopic completeness of the total $K_s\leq20.0$ sample
is 87\% and is one of the highest reached for samples of faint 
$K$-selected galaxies.

\section{Optical imaging and photometric redshifts}

The availability of high-quality and deep imaging (besides the $K_s$
images) and of spectroscopic redshifts covering also the critical
region of $1.3<z<1.9$ allowed us to optimize the estimate of
the photometric redshifts in a self-consistent way, and to
obtain the most reliable information possible on the redshifts of
the targets with no spectroscopic $z$ or with no observations
available.

Although the final multicolor data set is not homogeneous (in terms of
bandwidths and depth) over the two fields, the strategy followed in
both fields is identical, aimed at assuring that the UV and optical
coverage and depth are adequate to follow the spectral shape of the
reddest galaxies in the sample. In practice, we have imaged the 0055-269 
field over 10 bands ($UBGVR R_w IzJK_s$), obtained with the ESO
NTT + SUSI2 ($UBGV R_w I$) and SOFI ($J$), and VLT + FORS1 ($R$ and 
$z$), for a total of about 45 hours
of integration. In the CDFS field, we used a combination of
public EIS NTT data ($U$) and deep FORS1 images ($BV R Iz$) 
obtained in the framework of a different program (courtesy of
P. Rosati \& M. Nonino).
Unfortunately, the FORS1 images do not overlap
completely with the area surveyed here, so that multicolor imaging
and photometric redshifts are available for a subset of 272 over 348 
objects in the CDFS. In addition, and at variance with the 0055-269
field, the multicolor catalog of the CDFS is not in its final form, 
since we plan to eventually include the forthcoming $J$ EIS public images, 
but the current ($UBV R IzK$) data set already provides a highly
satisfactory solution for photometric redshifts.

For the purpose of estimating photometric redshifts and deriving
the Spectral Energy Distributions (SEDs) of the target galaxies,
we developed a technique to obtain a seeing--corrected ``optimal'' 
aperture photometry to cope with the inhomogeneous image quality 
over the different bands.
Photometric redshifts have been obtained through template fitting (see
Fontana et al 2000 et al. for the details), using in this case the
PEGASE 2.0 template spectral package (Fioc \& Rocca--Volmerange 1999).

The whole procedure has been tested against the sample with secure
spectroscopic redshift, and the results are shown in Fig. 9-10
The comparison between spectroscopic and photometric redshifts shows
that it is possible to achieve a high reliability and accuracy 
even with ground--based data sets, provided that
the imaging data are of adequate quality. 

The results in the two fields are slightly different, because of 
the different quality of the imaging data set. 
To compute a suitable statistics on $z_{spe}-z_{phot}$, one has to
define a recipe to account for the non-gaussian shape of its
distribution function. The one we adopt here is to compute mean
values and standard deviations of $z_{spe}-z_{phot}$ on all the
objects with a fractional error $\Delta z=(z_{spe}-z_{phot})/(1+z_{spe})$ 
smaller (in modulus) than $0.15$. This choice is based on the fact that 
the fractional error $\Delta z$ is roughly constant with redshift with a 
$\sigma$--clipped dispersion of about 0.05 in both these samples and 
in the HDFN (Cohen et al. 2000, Fontana et al 2002). The 
corresponding selection is delimited by the two dashed lines in Fig. 
9. Objects outside this range can be described as "outliers".
As it emerges from Fig. 9, most of the objects lie within the
boundaries that we use to define the ``satisfactory'' fits, and most
of the outliers are close to the same boundaries.

In the 0055-269 field, the number of outliers is fairly small (7 over
140), and the r.m.s. dispersion of the remaining fraction in the range
$0<z<2$ is $\sigma_{\Delta z}=0.078$, with a median value $<\Delta
z>=-0.004$, very close to the values obtained in the HDF--N over the
same redshift range and with the same template set (Fontana et al
2000). In the case of the CDFS, we reached a final accuracy of 
$\sigma_{\Delta z}=0.095$ with a median value $<\Delta z>=0.02$,
with a fraction of discrepant objects of 13 over the 198 galaxies
with secure redshifts used in the comparison.

It is also important to notice that objects with ``lower quality''
spectroscopic redshifts (see section 5) generally follow the
relation between photometric and spectroscopic redshifts (Fig. 9).

In both cases, the high accuracy and the low level of misidentified 
objects allow us to use the photometric redshift estimate both to 
support the redshift identification of objects with uncertain 
spectroscopic redshift, and to adopt the photometric redshift for the 
unobserved/unidentified objects.

The dispersion on the global sample (CDFS + 0055-269 fields) is 
$\sigma=0.089$ and the average is $<z_{spe}-z_{phot}>$=0.012.

In the whole K20 database, there are only 9 objects (in the CDFS)
for which neither $z_{spe}$ nor $z_{phot}$ are available. Thus, the
overall redshift completeness reached for our sample by using both 
$z_{spec}$ and $z_{phot}$ is 537/546=98.3\%.

\begin{figure}[ht]
\resizebox{\hsize}{!}{\includegraphics{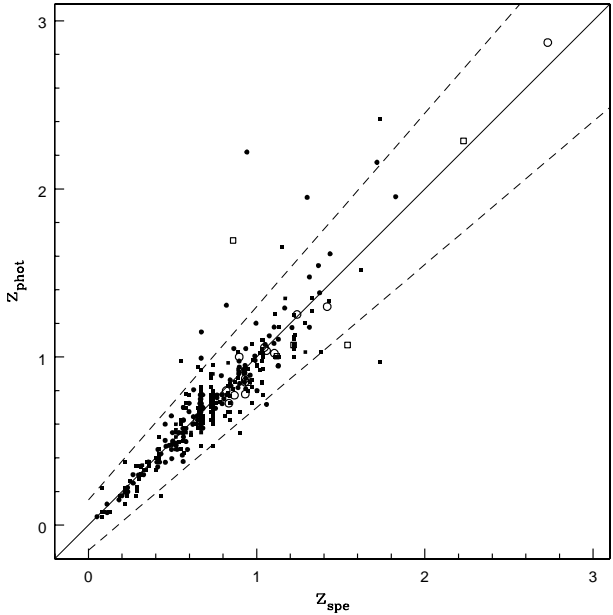}}
\caption{\footnotesize
The comparison between spectroscopic (x--axis) and photometric (y--axis)
redshifts of the galaxies in the K20 survey: 140 galaxies with secure 
redshifts in the 0055-269 field indicated by filled circles and 
198 galaxies in the CDFS indicated by filled squares. 
The solid line shows the $z_{spe}=z_{phot}$ relation. The two
dashed line delimit the region where $|z_{spe}-z_{phot}|\leq
0.15(1+z_{spe})$,that has been used to compute the overall statistics
in the sample (see text for details). The open symbols indicate objects
with ``lower quality'' spectroscopic redshifts.
}
\label{fig:plot}
\end{figure}

\begin{figure}[ht]
\resizebox{\hsize}{!}{\includegraphics{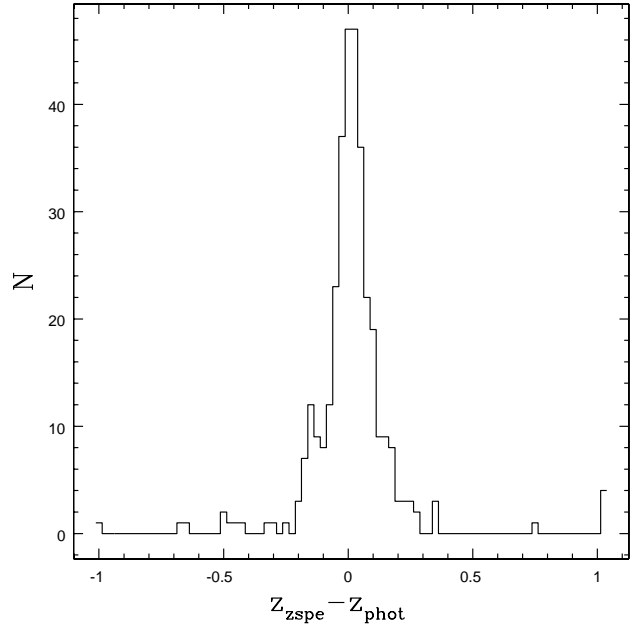}}
\caption{\footnotesize
The histogram of the observed scatter $z_{spe}-z_{phot}$ for all 
the galaxies of the sample with spectroscopic redshift. }
\label{fig:plot}
\end{figure}

\begin{table*}
\caption[]{Extract of the $K_s$-band photometric catalog of the 0055-269
field}
\begin{tabular}{lcccc} \hline\hline
& & & & \\
ID & Right Ascension (J2000) & Declination (J2000) & $K_s$ & $\sigma(K_s)$ \\
& & & & \\ \hline\hline
00013	& 00:57:55.27	& -26:40:58.3	& 18.83	& 0.03\\
00017	& 00:57:58.24	& -26:40:59.6	& 19.68	& 0.07\\
00018	& 00:58:02.90	& -26:41:01.5	& 16.16	& 0.01\\
00019	& 00:58:06.68	& -26:40:59.7	& 18.98	& 0.04\\
00021	& 00:57:51.54	& -26:41:03.6	& 18.54	& 0.03\\
00022	& 00:57:59.53	& -26:41:04.6	& 19.19	& 0.05\\
00023	& 00:57:56.82	& -26:41:05.2	& 19.87	& 0.08\\
00024	& 00:57:54.74	& -26:41:07.9	& 19.94	& 0.06\\
00025	& 00:57:54.52	& -26:41:09.7	& 18.46	& 0.02\\
.....	& ...........	& ........... 	& .....	& ....\\
\end{tabular}
\end{table*}

\section{Release of the $K_s$-band photometric catalog}

We make the $K_s$-band photometric sample available at the web site 
{\tt http://www.arcetri.astro.it/$\sim$k20/releases/}. This version 
of the sample includes for the moment only the names of the targets, 
the coordinates, the total (BEST) $K_s$-band magnitudes and their 
photometric uncertainties. An extract of such catalog is shown in
Table 5. In the next papers, we will also release the full spectroscopic 
and multicolor catalogs.

\section{Summary}

We discussed the general scientific background and aims of the
K20 survey and we described the sample photometric and spectroscopic 
properties. 

The relevant advantages of the K20 database are the small 
K-correction effects due to the $K_s$-band selection, its size 
(about 500 galaxies), the distribution of the targets in two 
independent fields, the use of near-IR spectroscopy for a 
subsample of the targets, and the availability of a large 
deep imaging database from the optical to the near-IR. 

Extensive simulations show that the sample is highly complete 
to $K_s=20.0$ and not affected by strong selection effects
on the galaxy population, with the possible exception of a 
slight underestimate of the total flux for large and luminous
early-type galaxies with de Vaucouleurs surface brightness 
profiles. Such a selection effect can be taken into account
when comparing the observed K20 sample with model predictions.
The observed galaxy counts and the $R-K_s$ color distribution
are in agreement with literature results. 

Optical spectroscopy, aided by near-IR spectroscopy, allowed us 
to achieve a high redshift completeness for a sample of galaxies 
selected in the near-IR (94\% to $K_s<19.0$ and 87\% $K_s<20.0$), 
and to obtain spectroscopic redshifts in the ``desert'' of 
$1.4<z<2.2$. 

The high-quality deep imaging database allowed us also to obtain 
tested and reliable photometric redshifts for the unobserved or 
spectroscopically unidentified galaxies. Using the photometric 
redshifts, the global completeness ($z_{spec}+z_{phot}$) increases 
to 98.3\% of the total sample.

Such a rich photometric and spectroscopic database makes the 
K20 sample a key tool to investigate the formation and evolution 
of galaxies.

\begin{acknowledgements}
We thank the VLT support astronomers for their kind assistance
and competent support
during the observations. AC warmly thanks ESO (Garching) for the 
hospitality during his visits. We thank J. Cohen for providing 
the CFGRS sample in digital form. We warmly thank Piero Rosati and Mario
Nonino for providing the reduced and calibrated $BVRI$ FORS1 images of the
CDFS. We also thank the anonymous referee for the useful comments.
The imaging observations of the 0055-269 field were performed 
during SUSI2 guaranteed time of the Observatory of Rome in the 
framework of the ESO-Rome Observatory agreement for this instrument.

\end{acknowledgements}


\begin{thebibliography}{}
\bibitem[]{}{Angeretti L., Pozzetti L., Zamorani G., 2002, in "A new era in
cosmology", Durham, UK, ASP Conference Series, ed. N.Metcalfe \&
T.Shanks, in press}
\bibitem[1996]{bcf}{Baugh C.M., Cole S., Frenk C.S., 1996, MNRAS 283, 1361}
\bibitem[1996]{bcf2}{Baugh C.M., Benson A.J., Cole S., Frenk C.S., Lacey
C., 2002, in `The Mass of Galaxies at Low and High Redshift', Venice
2001, eds. R. Bender, A. Renzini, in press, astro-ph/0203051}
\bibitem[]{}{Bender R., Burstein D., Faber S.M., 1992, ApJ, 399, 462}
\bibitem[]{}{Bertin E., Arnouts S. 1996, A\&A, 117, 393}
\bibitem[]{}{Broadhurst T.J., Ellis R.S., Glazebrook K. 1992, Nature,
355, 55}
\bibitem[]{}{Bruzual G., Charlot S. 1993, 405, 538}
\bibitem[]{}{Cimatti A. et al. 2002, A\&A, 381, L68}
\bibitem[]{}{Cohen J.G., Blandford R., Hogg D.W., Pahre M.A., Shopbell
P. 1999a, ApJ, 512, 30}
\bibitem[]{}{Cohen J.G., Hogg D.W., Pahre M.A., Blandford R., Shopbell
P., Richberg K. 1999b, ApJS, 120, 171}
\bibitem[]{}{Cohen J.G., Hogg D.W., Blandford R., Cowie L.L., Hu E.M.,
Songaila A., Shopbell P., Richberg K. 2000, ApJ, 538, 29}
\bibitem[]{}{Cole S., Lacey C.G., Baugh C.M., Frenk C.S. 2000, MNRAS,
319, 168}
\bibitem[]{}{Cole S. et al. 2001, MNRAS, 326, 255}
\bibitem[]{}{Cowie, L.L., Gardner J.P., Hu E.M., Songaila A., Hoddapp
K.-W., Wainscoat R.J. 1994, ApJ, 434, 114}
\bibitem[]{}{Cowie, L.L., Songaila A., Hu E.M., Cohen J.G. 1996, AJ,112,839}
\bibitem[2000]{daddi}{Daddi E., Cimatti A., Pozzetti L., et al., 2000, A\&A 361, 535}
\bibitem[2000]{daddi3}{Daddi E., Broadhurst T., Zamorani G., Cimatti A.,
R\"ottgering H.J.A., Renzini A., 2001, A\&A, 376,825}
\bibitem[]{}{Daddi E. et al. 2002, A\&A, 384, L1}
\bibitem[]{}{Dalcanton J.J., 1998, ApJ, 495, 251}
\bibitem[]{}{Djorgovski, S. et al., 1995, ApJ, 438, L13}
\bibitem[]{}{Drory N., Bender R., Snigula J., Feulner G., Hopp U., 
Maraston C., Hill G.J., de Oliveira C.M. 2001, ApJ, 562, L111}
\bibitem[]{}{Fasano G., Cristiani S., Arnouts S., Filippi M. 1998, AJ,
115, 1400}
\bibitem[]{}{Firth A.E., Somerville R.S., McMahon R.G. et al. 2001,
MNRAS, submitted (astro-ph/0108182)}
\bibitem{}{Fontana A., Menci N., D'Odorico S., Giallongo E., Poli F.,
Crisitiani S., Moorwood A., Saracco P. 1999, MNRAS, 310, L27}
\bibitem{}{Fontana A. et al. 2002, A\&A, submitted}
\bibitem[]{}{Fontana A., D'Odorico S., Poli F., Giallongo E., Arnouts S.,
Cristiani S., Moorwood A., Saracco P. 2000, AJ, 120, 2206}
\bibitem[]{}{Gardner, J. P.; Cowie, L. L.; Wainscoat, R. J., 1993,
ApJ, 415, L9}
\bibitem[]{}{Gardner, J. P.; Sharples, R. M.; Carrasco, B. E.; Frenk, C. S.,
1996, MNRAS, 282, L1}
\bibitem{}{Gavazzi G., Pierini D., Boselli A. 1996, A\&A, 312, 397}
\bibitem[]{}{Gehrels N. 1986, ApJ, 303, 336}
\bibitem[]{}{Giacconi R., Rosati P., Tozzi P. et al. 2001, ApJ,551,624}
\bibitem[]{}{Glazebrook, K.; Peacock, J. A.; Collins, C. A.; Miller, L.,
1994, MNRAS, 266, 65}
\bibitem[]{}{Hall P.B., Green R.F., Cohen M. 1998, ApJS, 119, 1}
\bibitem[]{}{Huang, J.-S.; Cowie, L. L.; Gardner, J. P.; Hu, E. M.; 
Songaila, A.; Wainscoat, R. J., 1997, ApJ, 476, 12}
\bibitem[]{}{Impey C. D., Sprayberry D., Irwin M. J., Bothun G. D, 
1996, ApJS, 105, 209}
\bibitem[]{}{Impey C. D., Bothun G. D, 1997, ARA\&A, 35, 267}
\bibitem[1996]{kau93}{Kauffmann G., White S.D.M., Guiderdoni B. 1993, MNRAS, 
264, 201}
\bibitem[1996]{kau96}{Kauffmann G., 1996, MNRAS, 281, 487}
\bibitem[1996]{kau96}{Kauffmann G., Charlot S. 1998, MNRAS, 297, L23}
\bibitem[]{}{Kinney A.L., Calzetti D., Bohlin R.C. et al. 1996, ApJ,467,38}
\bibitem[]{}{Kochanek C.S., Pahre M.A.,Falco E.E. et al. 2001,ApJ,560,566}
\bibitem[]{}{Kron R.G., 1980, ApJS, 43, 305}
\bibitem[]{}{K\"ummel, M. W.; Wagner, S. J., 2000, A\&A, 353, 867}
\bibitem[]{}{Lilly S.J., Le Fevre O., Crampton D., Hammer F., Tresse L., 1995, ApJ, 455, 50}
\bibitem[]{}{Madau P., Pozzetti L., Dickinson M., 1998, ApJ, 498, 106}
\bibitem[]{}{Maihara T. et al., 2001, PASJ, 53, 25}
\bibitem[]{}{Marleau F.R., Simard L. 1998, ApJ, 507, 585}
\bibitem[]{}{Martini P., 2001, AJ, 121, 598}
\bibitem[2001]{}{McCarthy et al. 2001, ApJL, 560, L131}
\bibitem[2001]{}{McLeod, B. A.; Bernstein, G. M.;
 Rieke, M. J.; Tollestrup, E. V.; Fazio, G. G., 1995, ApJS, 96, 117}
\bibitem[2001]{}{Minezaki, T., Kobayashi Y., Yoshii Y., Peterson B.A.,
1998, ApJ, 494, 111}
\bibitem[]{}{Mobasher, B.; Ellis, R. S.; Sharples, R. M., 1986, MNRAS, 223, 11}
\bibitem[]{}{Moorwood A.F.M., Cuby J.-G., Lidman C. 1998, The Messenger,
91,9}
\bibitem[]{}{Moorwood A.F.M. et al. 1999, The Messenger, 95, 1}
\bibitem[]{}{Moustakas, L.A.; Davis, M., Graham, J.R., 
Silk, J., Peterson, B.A.; Yoshii, Y., 1997, 475, 445}
\bibitem[]{}{Pahre M.A., 1999, ApJS, 124, 127}
\bibitem{}{Peebles P.J.E. 2002, in ``A new era in cosmology'', Durham,
U.K., ASP Conference Series, ed. N. Metcalfe \& T. Shanks, in
press,astro-ph/0201015}
\bibitem{}{Persson S.E., Murphy D.C., Krzeminski W., Roth M., Rieke M.J.
1998, AJ, 116, 2475}
\bibitem{}{Rengelink R. et al. 1998, submitted to A\&A, astro-ph/9812190}
\bibitem[2001]{}{Renzini A. 1999, in ``The formation of galactic
bulges'', edited by C.M. Carollo, H.C. Ferguson, R.F.G. Wyse. Cambridge,
U.K., Cambridge University Press, p.9}
\bibitem[2001]{}{Renzini A., Cimatti A. 1999, in ``The Hy-Redshift
Universe: Galaxy Formation and Evolution at High Redshift'', 
Berkeley, USA, ASP Conference Proceedings, Vol. 193, Edited by Andrew J.
Bunker and Wil J. M. van Breugel, p. 312}
\bibitem[]{}{Rudnick G., Franx M., Rix H.-W., Moorwood A., Kuijken K.,
van Starkenburg L., van der Werf P., R\"ottgering H.J.A., van Dokkum P.,
Labbe I. 2001, AJ, 122, 2205}
\bibitem[]{}{Saracco, P.; Iovino, A.; Garilli, B.; Maccagni, D.; Chincarini, 
G., 1997, AJ, 114, 887}
\bibitem[]{}{Saracco P., D'Odorico S., Moorwood A., Buzzoni A., Cuby
J.-G., Lidman C. 1999, A\&A, 349, 751}
\bibitem[]{}{Saracco, P.; Giallongo, E.; Cristiani, S.; D'Odorico, S.; 
Fontana, A.; Iovino, A.; Poli, F.; Vanzella, E., 2001, A\&A, 375, 1}
\bibitem[]{}{Scalo J.M. 1986, Fundam. Cosm. Phys., 11, 1}
\bibitem[]{}{Schlegel D.J., Finkbeiner D.P., Davis M. 1998, ApJ, 500, 525}
\bibitem[]{}{Soifer, B. T. et al., 1994, ApJ, 420, L1}
\bibitem[]{}{Somerville R.S, Primack J.R.,Faber S.M. 2001, MNRAS, 320, 504}
\bibitem[]{}{Songaila A., Cowie, L.L., Hu E.M., Gardner J.P. 1994, ApJS,
94, 461}
\bibitem[]{}{Stern D., Connolly A., Eisenhardt P., Elston R., Holden B.,
Rosati P., Stanford A., Spinrad H., Tozzi P., Wu K. 2001, in ``Deep
Fields'', Proceedings of the ESO Workshop, Garching, Germany, 
Springer-Verlag, p. 76}
\bibitem[]{}{Szokoly, G.P.; Subbarao, M.U.; Connolly, A.J.; Mobasher, 
B., 1998, ApJ, 492, 452}
\bibitem[]{}{van Dokkum P.G. \& Stanford S.A. 2001, ApJL,562,L35} 
\end{thebibliography}
\end{document}